\documentclass[journal,transmag]{IEEEtran}

\usepackage{graphicx}
\usepackage{cite}
\usepackage{amsmath}
\usepackage{url}
\usepackage{float}

\begin{document}

\title{Magnetic Nanohorns for measurement of the interfacial Dzyaloshinskii-Moriya interaction: A micromagnetic study}

\author{\IEEEauthorblockN{
Thomas Wong\IEEEauthorrefmark{1}, and Vincent Sokalski\IEEEauthorrefmark{2}}

\IEEEauthorblockA{\IEEEauthorrefmark{1}Department of Physics,
Carnegie Mellon University, Pittsburgh, PA 15213 USA\\
\IEEEauthorrefmark{2}Department of Materials Science and Engineering,
Carnegie Mellon University, Pittsburgh, PA 15213 USA}

\thanks{
Corresponding author: Vincent Sokalski (email: sokalski@cmu.edu).}}


\IEEEtitleabstractindextext{%
\begin{abstract}
We evaluate the micromagnetic energy of a domain wall contained in a thin, horn-shaped nanostructure and propose a method to extract the strength of the interfacial Dzyaloshinskii-Moriya interaction based on the equilibrium position of the wall.  An in-plane magnetic field coupled with the interfacial DMI balances a competing driving force related to the “nanohorn” tapering and consequent variation in length.  It is evident that the strength of the the DMI is directly correlated to the equilibrium position of the DW.  Moreover, the range of measurable DMI values can be controlled by varying the degree of tapering in the nanohorn.  Numerical and semi-analytical calculations were separately performed and show broad agreement.  The calculations provided here offer an alternative framework for future measurement of DMI, which has been notoriously difficult to do experimentally.
\end{abstract}

}

\maketitle

\IEEEdisplaynontitleabstractindextext

\IEEEpeerreviewmaketitle

{\large\textbf{This work has been submitted to the IEEE for possible publication. Copyright may be transferred without notice, after which this version may no longer be accessible.}}

\section{Introduction}

\IEEEPARstart{T}{he} Dzyaloshinskii-Moriya interaction (DMI) is a critical property of magnetic thin films for the stabilization of chiral magnetic objects, including skyrmions, for use in future spintronic applications.\cite{Dzyaloshinsky1958, Moriya1960,muhlbauer2009,Yu2010,Woo2016} For the interfacial DMI, chiral N\'eel-type domain walls (DWs) are preferred, with the sign of DMI determining the direction of the domain wall internal magnetization.\cite{Thiaville2012} This chirality is critical in governing the efficiency with which both domain walls and skyrmions can be manipulated by spin currents.\cite{Emori2013, ryu2013}. However, direct measurement of the strength of this interaction remains an experimental challenge. Commonly used techniques include Brillioun Light Scattering (BLS) and asymmetric domain expansion, which have offered agreement only in some materials systems.\cite{arora2020variation,Hrabec2014,Lau2016,Lavrijsen2015,Lau2018}

Here, we propose the use of a thin film patterned into the shape of a "nanohorn" that would enable the interfacial DMI to be read directly from the equilibrium position of a magnetic domain wall.  The concept is based on a balance between two energetic driving forces acting on the domain wall.  First, the nanohorn tapering drives the DW towards the narrow end due simply to the favorable reduction in total length.  Second, an in-plane field coupled with the interfacial DMI will drive the DW toward the opposite end of the nanohorn depending on the sign of DMI.  This is due to the Zeeman energy asssociated with the in-plane field and the internal magnetization of the DW itself.  The result is an intermediate position that corresponds to the strength of the interfacial DMI, which we elaborate on below.

\section{Magnetic Nanohorn Design and Modeling}

\begin{figure}[h!]
\includegraphics[width = 3.5 in]{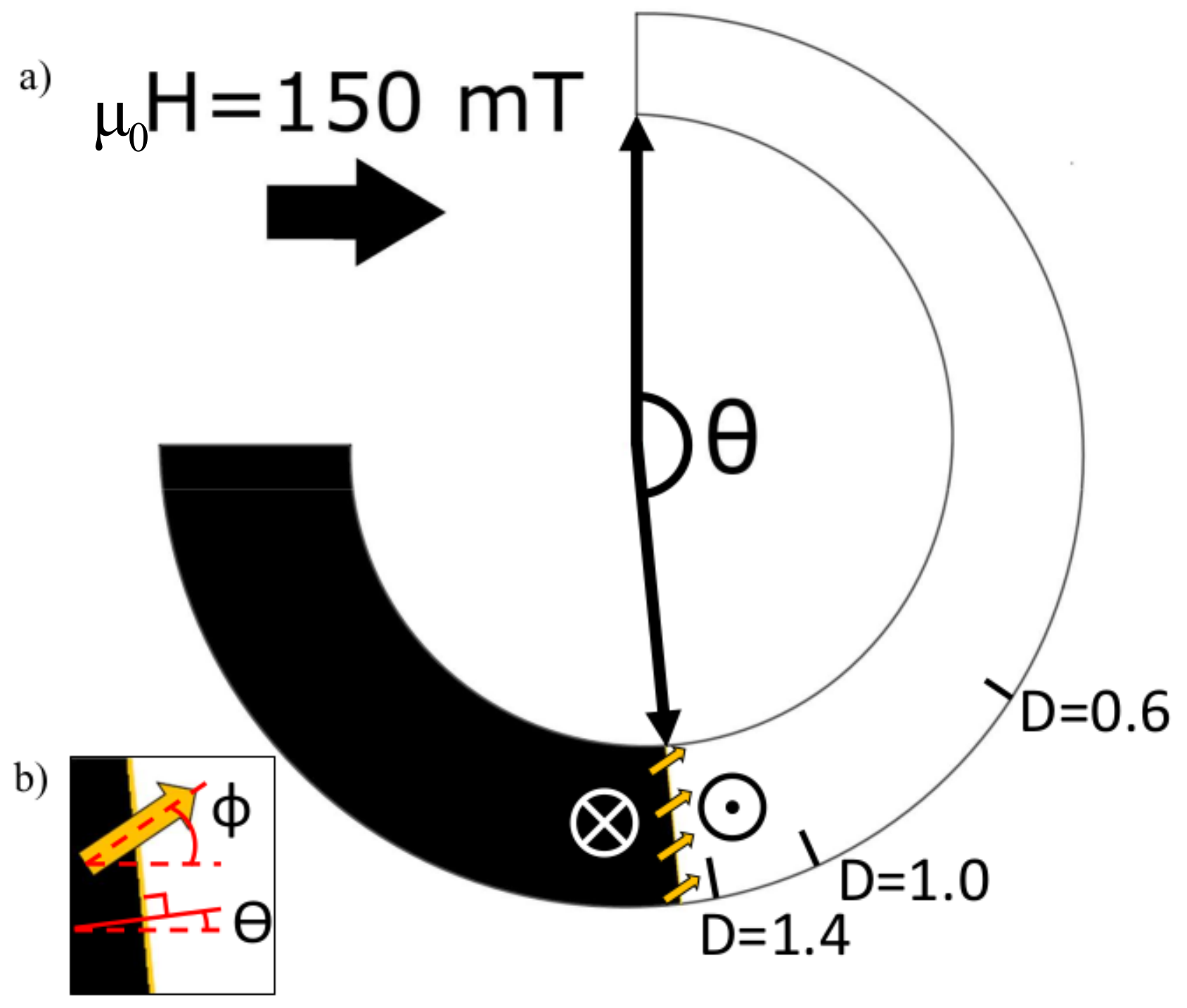}
\caption{\label{Figure1} Schematic of an example magnetic nanohorn with $c/b = 0.1$.  Tick marks indicate the equilibrium position of a DW for different values of DMI.  Lower left inset: coordinate system used for equation 2.}
\end{figure}

The proposed nanohorn design is shown in Fig \ref{Figure1}, which is defined as the area between two curves given by the following polar equations:

\begin{equation}
r(\theta)=a\pm b\mp c(\frac{3\pi}{2}-\theta)
\end{equation}

for $\theta$ ranging from $0$ to $\frac{3\pi}{2}$ (as defined in Fig \ref{Figure1}). This produces two spirals centered at a circle of radius $a$. The wider end of the horn has length $2b$ and the radial cross section of the horn decreases with slope $2c$.

\begin{figure}[h!]
\includegraphics[width = 3.5 in]{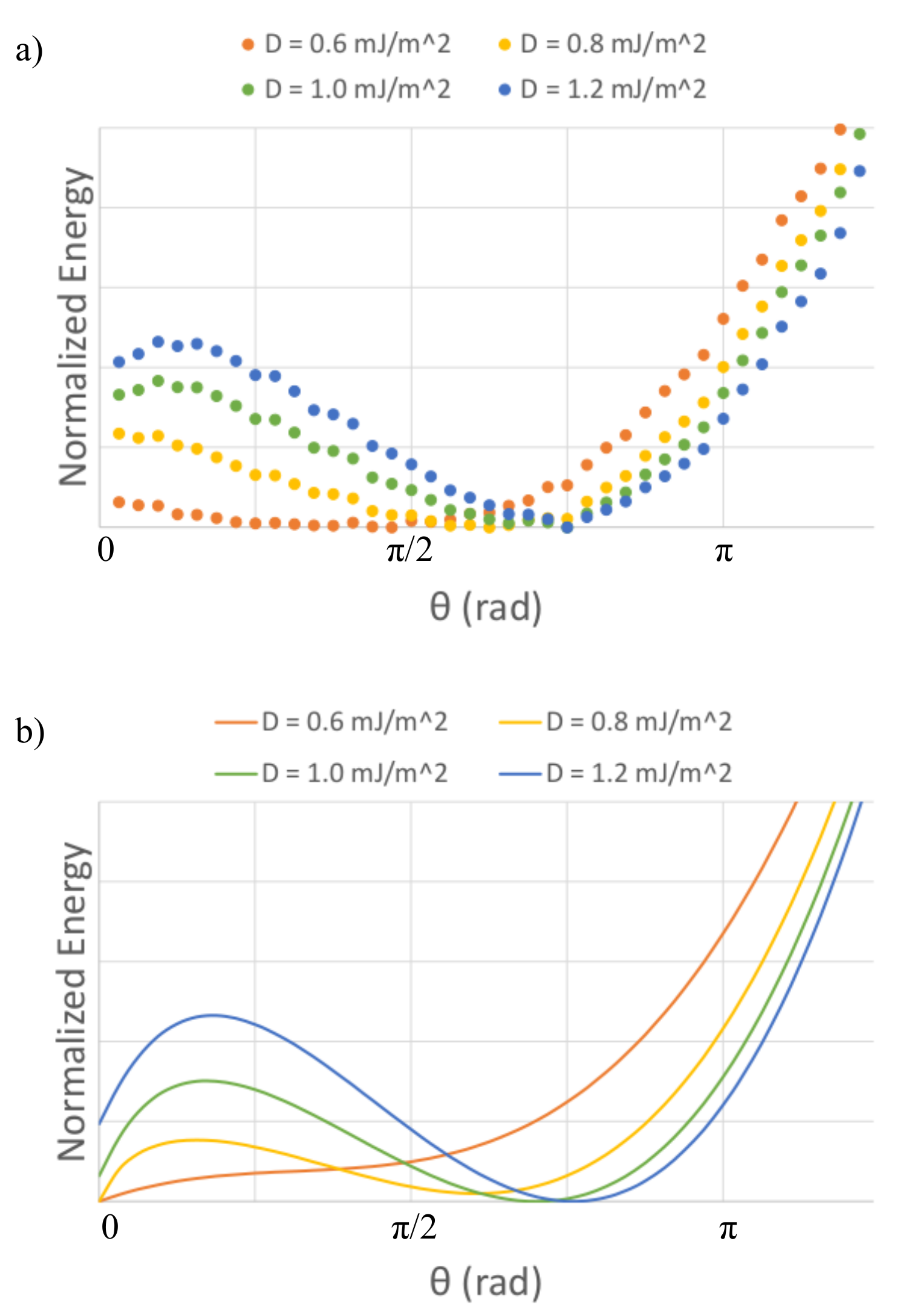}
\caption{\label{Figure2} The relaxed energy vs $\theta$ in both a) micromagnetic simulation and b) semi-analytical computation shows a minimum which changes position based on DMI.}
\end{figure}

\begin{figure}[h!]
\includegraphics[width = 3.5 in]{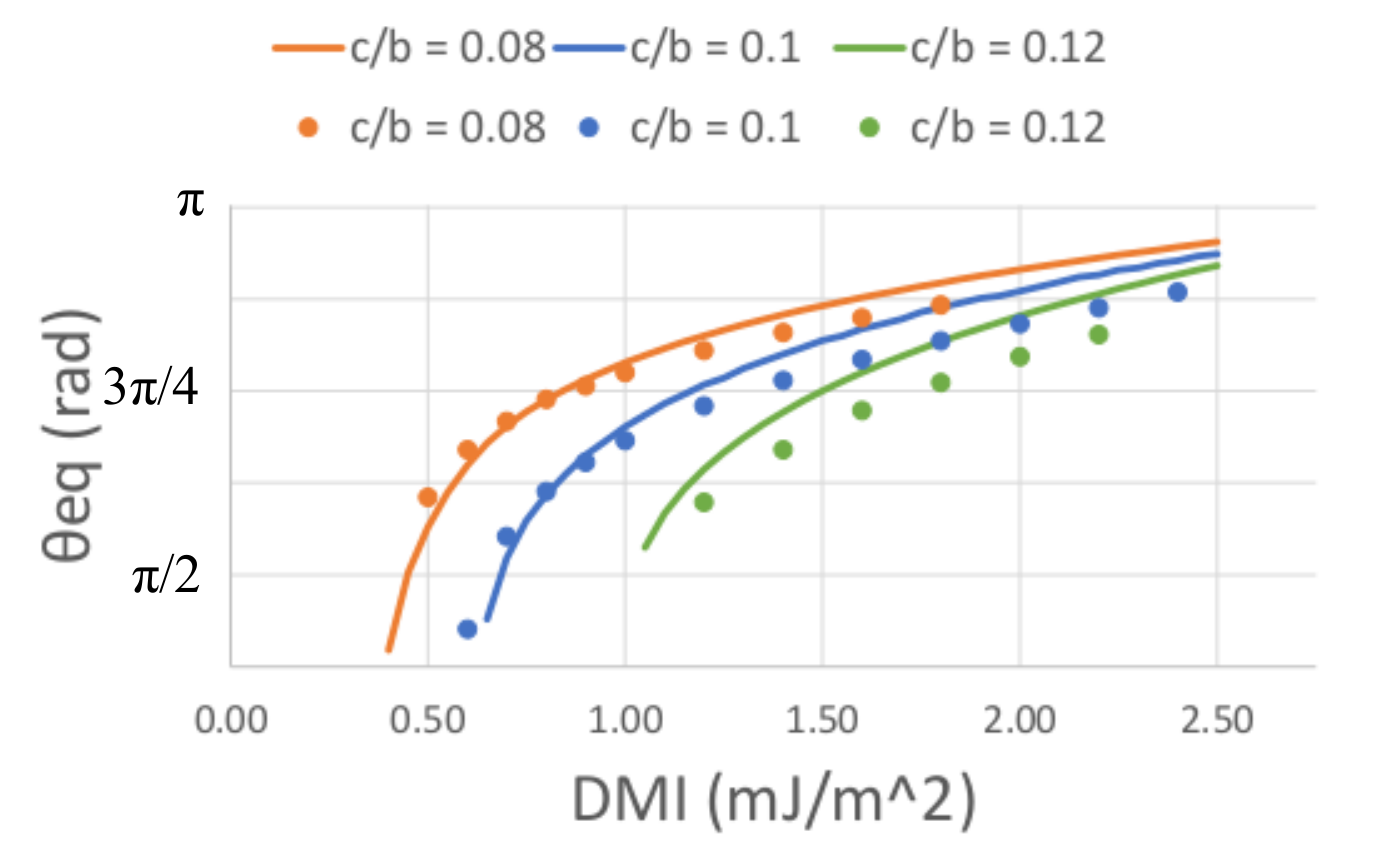}
\caption{\label{Figure3} In both micromagnetic simulation (dots) and semi-analytical computation (curves) for multiple horn shapes, equilibrium $\theta$ has a clear relationship with DMI and the $c/b$ tapering parameter.}
\end{figure}

A horn with shape parameters $(a,b,c)=(400 nm,100 nm,10 nm/rad)$ was used to perform micromagnetic energy calculations on a 1nm thin film.  We use the steepest conjugate gradient method built into the software package Mumax$^3$.\cite{vansteenkiste2014design} The cell size was fixed at 1x1x1 nm with the following material properties $M_s=800 kA/m$, $A=15 pJ/m$, and $K_{u}=1\times10^7 J/m^3$. We apply an in-plane field $\mu_{0}H_x=150mT$, which as noted previously, creates a Zeeman energy with the DW internal magnetization. A domain wall was positioned at a range of angles from $\theta=0$ to $\frac{3\pi}{2}$ and at each one, magnetization was relaxed to minimize total energy. The equilibrium position was determined from the minimum in a plot of relaxed energy vs $\theta$ (e.g. Fig \ref{Figure2}). The resulting equilibrium positions as a function of $D_{int}$ are shown in Fig \ref{Figure3} for two possible horn dimensions including both micromagnetic and analytical results.  It is notable that there is a finite range of $D_{int}$ that results in an equilibrium position that does not fall to one of the limits, which is discussed later.

We now turn towards a semi-analytical treatment of the DW position in the nanohorn. It is well-established that the energy of a magnetic domain wall in the thin film limit can be approximated by the following:\cite{Pellegren2017, Thiaville2012}

\begin{equation}
\begin{split}
    \sigma(\Theta,\phi) = \sigma_0 -\pi D_{int}\cos(\phi-\Theta) - \pi\lambda\mu_0H_xM_s\cos(\phi)\\ +\frac{ln(2)}{\pi}t_f\mu_0M_s^2\cos^2(\phi-\Theta)
\end{split}
\end{equation}

where $\Theta$ and $\phi$ represent the angles of the DW normal and internal magnetization relative to the applied field $H_x$, respectively. $D_{int}$ is the interfacial DMI and $M_s$ is the saturation magnetization. $\sigma_0 = 4\sqrt{AK_{eff}}$ is the Bloch wall energy and $\lambda = \sqrt{A/K_{eff}}$ is the Bloch width, where $A$ is the exchange stiffness and $K_{eff} = K_{u} - \frac{\mu_0}{2}M_s^2$ is the effective anisotropy constant. The length of the domain wall as a function of $\Theta$ in the nanohorn can be derived:

\begin{equation}
\begin{split}
l(\Theta)=2b-2c(\Theta+\frac{\pi}{2})
\end{split}
\end{equation}

Here we ignore possible DW tilting, which is generally consistent with the results of the micromagnetic modeling. In this case, $\theta=\pi-\Theta$.

To compare with the Mumax3 simulations, we used the same horn shape and material parameters. Note that $b$ and $c$ carry the same scale factor and $\frac{c}{b}$ is the parameter which defines tapering. Given equations for DW energy per unit volume and length, (2) and (3), we consider their product the DW energy per unit length. At each particular value of $\Theta$, the DW energy per unit area was minimized with respect to DW internal magnetization $\phi$ to determine the relaxed energy as a function of $\Theta$. As with the micromagnetic calculations, the equilibrium position of the DW is determined from the minimum in energy vs $\Theta$. We repeat this for a range of DMI and compare with the date from Mumax3 simulation, which is superimposed in Fig \ref{Figure3}.

\section{Discussion}

In both the micromagnetic simulations and semi-analytical calculations we show a nonlinear relationship between DMI and equilibrium DW position. Low DMI below a particular threshold is too weak to result in an equilibrium position and instead the domain wall vanishes at the shorter end of the nanohorn. It is noteworthy that this critical value can be shifted to higher values of DMI by increasing the degree of tapering. Above that threshold DMI, there is a steep change in $\theta_{eq}$, which levels off and asymptotically approaches $\pi$. This is a function of the slope of the nanohorn and can be tuned to either reveal equilibria for lower DMI by decreasing the degree of tapering or be more sensitive at higher DMI by increasing the tapering.

For larger nanohorns, there is also the possibility of the domain wall no longer remaining radial with the center of the horn. In other words, the domain wall can tilt, introducing further complexity, which may be the origin of some of the disagreement between the micromagnetic and analytical results. In an experimental context, these nanohorns could be patterned optically and imaged by Kerr microscopy allowing for the simultaneous examination of a multitude of devices.  To account for uncertainty in the initial guess of DMI, it would be reasonable to create an array of devices with varying dimensions to improve the precision of the measured DMI.  We also acknowledge that there may be limitations in driving the wall to its equilibrium. This could be addressed through increased statistics (simultaneously examining many nanohorns of different dimensions) or exciting the DW with AC magnetic fields to overcome edge pinning.

\section{Conclusion}

We present computational and semi-analytical evidence for the use of magnetic nanohorns to measure interfacial DMI. The combination of a tapered shape and applied in-plane field creates an equilibrium position for a domain wall. We show that this equilibrium position varies with DMI, which may provide a technique for determining DMI of a thin film.

\section*{Acknowledgments}

The authors acknowledge valuable discussions with Michael Kitcher, Maxwell Li, and Nisrit Pandey. The authors acknowledge use of computational facilities of the Materials Characterization Facility at CMU under grant \# MCF-677785.

\ifCLASSOPTIONcaptionsoff
  \newpage
\fi

\bibliographystyle{IEEEtran}
\bibliography{main}



\end{document}